\documentclass[12pt]{article}
\usepackage{amsmath}
\usepackage{amstext}
\usepackage{amsfonts}
\usepackage{graphicx,epsfig}
\usepackage{indentfirst}
\usepackage{setspace}

\begin{document}

\title{ New Possibilities for Testing Local Realism in
High Energy Physics}

\author{Junli Li$^{a}$\; ~and
Cong-Feng Qiao$^{a,b}$\footnote{Email: qiaocf@gucas.ac.cn; Tel: +86-10-88256284}\\[0.5cm]
{\small $a)$ College of Physical Sciences, Graduate
University of Chinese Academy of Sciences}  \\
{\small YuQuan Road 19A, 100049, Beijing, China}\\
{\small $b)$ Theoretical Physics Center for Science Facilities
(TPCSF), CAS}\\
}

\date{}

\maketitle

\begin{abstract}
The three photons from the dominant ortho-positronium decay and two
vector mesons from the $\eta_c$ exclusive decays are found to be in
tripartite and high-dimensional entangled states, respectively.
These two classes of entangled states possess the Hardy type
nonlocality and allow a priori for quantum mechanics vs local
realism test via Bell inequalities. The experimental realizations
are shown to be feasible, and a concrete scheme to fulfill the test
in experiment via two-vector-meson entangled state is proposed.
\end{abstract}

\vspace{2cm}

\begin{minipage}{14cm}
\hspace{3mm} {\bf Key words:} High Energy Process, Entanglement,
Quantum Nonlocality
\end{minipage}

\vspace{7mm}

Bell's theorem \cite{bell-theorem} states that no local hidden
variable theory (LHVT) can reproduce all the possible results of
quantum mechanics (QM). This can be testified via either
inequalities, the Bell inequalities (BIs), satisfied by the
expectation values of LHVT while violated by QM predictions, or
logic contradictions remaining in LHVT vs QM. The original BI and
its generalizations mainly concern about two-qubit systems and have
been verified with various physical systems, see, e.g.,
\cite{clauser-rpp, bertlmann, genovese-pr}. However, the analogous
tests by virtue of high-dimensional and multipartite entangled
states, especially in high energy experiment are very limited
\cite{bertlmann,bell-ie}.

In late 1980s, Greenberger, Horne, and Zeilinger (GHZ) showed that
certain three(or more)-partite entangled states may exhibit
nonclassical effects more prominently than by bipartite state
\cite{ghz}. Soon afterwards, Hardy \cite{Hardy-nonmaximal} invented
a cute method to demonstrate the essential contradiction between QM
and LHVT without employing the BI, in which the non-maximally
entangled state works. A three-photon GHZ state was obtained in
\cite{3photon} by manipulating the parametric downconversion (PDC)
photons and the experimental result deviates the local realistic
expectation by more than eight standard deviations
\cite{3photon-nonlocal}. For the Hardy's type of violation, it was
proved that the argument may take the form of Clauser and Horne
\cite{Clauser-Horne} (CH) inequality and then reduce the required
detector efficiencies for loophole-free tests of Bell inequalities
\cite{Eberhard-ineq,hardy-effi}. Experimental measurement of the
Hardy fraction was obtained in \cite{non-pr-ch-ut}. It is noteworthy
that the existing entanglement tests are performed mainly in the
regime of visible light. The non-maximally entangled kaon pairs were
once suggested to realize the Hardy's type of criterion in
experiment \cite{hardy-kaon-1,hardy-kaon-2,hardy-kaon-3}, however
there has been no such experiment carried out yet to the best of our
knowledge.

In this Letter we propose to investigate the quantum non-local
entanglement in two well-established high energy processes, in which
the multipartite or/and high-dimensional entanglement is naturally
realized, i.e., the three photons in the dominant ortho-positronium
decay and the two vector mesons in the $\eta_c$ exclusive decay. In
\cite{aacin}, Acin, Latorre and Pascual already realized that the
ortho-positronium decay can provide an entangled state of three
space-separated photons. However, in this work we find that there is
a Hardy type contradiction between QM and LHVT, and the CH type
inequality would be violated by the quantum correlation of this
state.

In the middle of the last century, the polarization correlation of
the two photons in para-positronium decay was used to determine its
parity \cite{wheeler, exp-wu, cny}. In quantum mechanics these two
correlated photons are in the state of \cite{AB}
\begin{eqnarray}
|\Psi\rangle = \frac{1}{\sqrt{2}} \{ |x\rangle|y\rangle -
|y\rangle|x\rangle \}\; ,
\end{eqnarray}
where $|x\rangle$ and $|y\rangle$ stand for the photon polarization
states in coordinate space. The ortho-positronium, the spin triplet
state of electron and positron with $\rm{J^{PC}} = 1^{--}$, decays
overwhelmingly to three photons. In this case, the photon
polarization information can be extracted from the decay amplitude,
which is proportional to \cite{zuber}
\begin{eqnarray}
V \propto \sum_{\rm{cyc}} [ \epsilon_1 (\epsilon_2\cdot\epsilon_3 -
\delta_2\cdot\delta_3) + \delta_1(\epsilon_3\cdot\delta_2 +
\epsilon_2\cdot\delta_3) ]\; ,
\end{eqnarray}
where $\epsilon_{i}$ denote the photon polarization vectors, $k_i$
represent their momenta, and $\delta_i = k_i \times \epsilon_i$. In
the form of circular polarization, $\epsilon^{\pm}_i = \epsilon_i
\pm i\delta_i$ and the amplitude reads
\begin{eqnarray}
V \propto \sum_{{\rm cyc}} [
\epsilon_1^{+}(\epsilon_2^{-}\cdot\epsilon_3^{-}) +
\epsilon_1^{-}(\epsilon_2^{+}\cdot\epsilon_3^{+}) ]\; .
\end{eqnarray}
Obviously, $\epsilon_{i}^{\pm}$ represent the left- and right-handed
polarizations of photons, respectively. Thus the three photon
polarization state should be in the following form
\begin{eqnarray}
|3\gamma\rangle & = & \frac{1}{\sqrt{6}} \{ |RRL\rangle +
|RLR\rangle + |LRR\rangle \nonumber \\ & & \hspace{0.31cm}{} +
|LLR\rangle + |LRL\rangle + |RLL\rangle \}\; . \label{helic-b}
\end{eqnarray}
Here, $L$ and $R$ denote the left and right circularly polarized
photons, respectively.

As shown in \cite{three-qubit}, under stochastic local operations
and classical communication (SLOCC) pure three qubits entangled
states have six inequivalent classes, and among them two are
genuinely entangled, named as GHZ and W states. By virtue of the
method proposed in \cite{2nn}, one can easily confirm that the state
(\ref{helic-b}) is SLOCC equivalent to GHZ state. On the other hand,
from the point of entanglement measure, 3-tangle $\tau$ defined in
\cite{Distr-ent}, the three-qubit entangled state (\ref{helic-b})
has $ \tau(|3\gamma\rangle) = 1/3 $. Since $\tau$ is none zero only
in the case of GHZ or its equivalent states \cite{three-qubit}, we
can also conclude that (\ref{helic-b}) is in SLOCC equivalence to
the GHZ state. Therefore, we know that the three photons from the
ortho-positronium decay constitute the simplest multipartite
entangled system, the three-qubit state.

To observe the non-locality of state (\ref{helic-b}), one needs also
to measure the linear polarization of the state, e.g., the
horizontal and vertical polarizations $H$ and $V$. The
transformation between circular and linear polarizations reads
\begin{eqnarray}
\left(
\begin{array}{c}
    R \\
    L \\
\end{array}
\right) =
\frac{1}{\sqrt{2}}\left(
  \begin{array}{cc}
    1 & i \\
    1 & -i \\
  \end{array}
\right) \left(
  \begin{array}{c}
    H \\
    V \\
  \end{array}
\right)\; .
\end{eqnarray}
Under the linear polarization bases, the three-photon polarization
wave function can be reexpressed as
\begin{equation}
|3\gamma\rangle = \frac{1}{\sqrt{12}} \{ 3 |HHH\rangle + |HVV\rangle
+ |VHV\rangle + |VVH\rangle \} \; . \label{polar-b}
\end{equation}

To show the incompatibility of quantum mechanics with local
realism(LR), we need an additional form of the state
(\ref{helic-b}). Here, we express one photon in linear polarization,
but left the other two in circular bases. In this case the wave
function is
\begin{eqnarray}
|3\gamma\rangle &=& \frac{1}{\sqrt{12}} \{ (|RR\rangle + |LL\rangle
+ 2|RL\rangle + 2|LR\rangle) |H\rangle \nonumber\\
&&\hspace{0.45cm}{} - i ( |RR\rangle - |LL\rangle) |V\rangle \}\;
.\label{mix-b}
\end{eqnarray}

From Eqs.(\ref{helic-b}), (\ref{polar-b}) and (\ref{mix-b}) one can
easily get the following probabilities:
\begin{eqnarray}
P_{|3\gamma\rangle}(i=V,j=V) =  \frac{1}{4}\; ,\; \label{contr-1}
P_{|3\gamma\rangle}(C_{j}=C_{k}|\,i=V)  =  1\; ,~~~\label{contr-2}
\nonumber\\ P_{|3\gamma\rangle}(C_{i}=C_{k}|\,j=V)  =  1\; ,\;
\label{contr-3} P_{|3\gamma\rangle}(C_{i}=C_{j}=C_{k})  = 0\; .~~~
\label{contr-4}
\end{eqnarray}
Here, $P_{|3\gamma\rangle}(i=V,j=V)$ denotes the probability of two
of the photons being vertically polarized; $P_{|3\gamma\rangle}
(C_{j} = C_{k}| \,i=V)$ denotes the conditional probability of
photons $j$ and $k$ possessing the same circular polarization while
the photon $i$ being vertically polarized; and
$P_{|3\gamma\rangle}(C_{i}=C_{j}=C_{k})$ represents the probability
of all three photons having the same circular polarization.

The probabilities in (\ref{contr-4}) can be embedded into CH type
inequality \cite{cabello-ghz-w}, that is
\begin{eqnarray}
&& \hspace{0.28cm} P_{|3\gamma\rangle}(i=V, j=V)  -
P_{|3\gamma\rangle}( i=V, C_j\neq C_k)  \nonumber \\ && -
P_{|3\gamma\rangle}(C_i \neq C_k, j=V)   - P_{|3\gamma\rangle}(C_i =
C_j = C_k) \leq 0\; .\nonumber \\  \label{ch-bell}
\end{eqnarray}
This inequality is a constraint imposed by the LHVT. While inputting
the quantum mechanics results on $P_{|3\gamma\rangle}$, i.e., Eq.
(\ref{contr-4}), into (\ref{ch-bell}), one readily finds that
\begin{eqnarray}
\frac{1}{4} -0 -0 -0 \leq 0 \; ,
\end{eqnarray}
which is obviously a contradiction. Note that inequality
(\ref{ch-bell}) is valid only in the ideal case of perfect detection
efficiencies, by which only a restricted class of Local Realistic
Theories can be tested.

To perform the experimental test on above inequality becomes
realistic in recent years. Intense pencil beams of ortho-positronium
($30$\,-$1.7 \times 10^{4}s^{-1}$) can be produced based on a low
energy positron storage ring equipped with an electron cooling
system \cite{ps-beam}. This may accumulate about $10^{12}$
three-photon events in ortho-positronium decay a year. Unlike in the
case of Bell inequality where a group of measurements with relative
angles in the measurement bases should be performed, here only two
kinds of polarization measurements are necessary. However, we should
point out that the measurement of photon polarization in positronium
decay is a bit challenging since the photon energy lies in the
regime of hard X-ray. The entangled photons from ortho-positronium
decay are of hundreds of keV which is about 5 orders greater than
ordinary photons in visible region. In experiment, the X-ray
polarimeter has made the measurements on linear and circular
polarization components feasible in a wide range of energy. With
ordinary inorganic crystal GaAs, people had already realized the
measurement on polarizations of photon with energy span of 3-20 keV
\cite{x-ray-crys}. Recently, a novel kind of X-ray polarimeter based
on hexagonal crystal, which can separate different polarization
components simultaneously, is available \cite{x-ray-polar}. Although
its working energy is a bit lower than that of photon energy from
positronium decay, further development on X-ray polarization
analyzer is quite realistic in near future.

\begin{figure}[m]
\centering
\includegraphics[width=4.5cm,height=3.5cm]{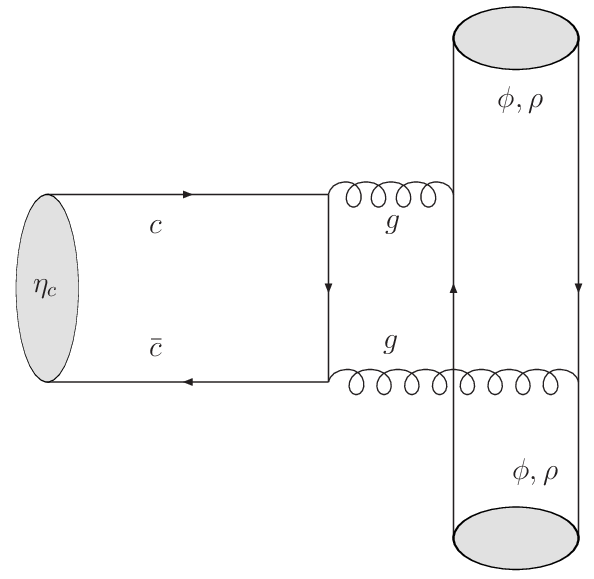}
\caption{}
\label{etacvv}
\end{figure}

As mentioned in the beginning, we find that a $3\times3$ entangled
state in spin can be produced in the process of $\eta_c$ exclusive
decay into two light vector mesons, i.e., $\eta_c \rightarrow
\rho\rho, \phi\phi$, etc., as shown in Figure \ref{etacvv}. In the
following part we briefly prove this, show its nature, and propose a
scheme on how to carry out the test on LR by using of the
two-vector-meson entangled state. Details will be presented
elsewhere.

\begin{figure}[t]
\centering
\includegraphics[width=6cm,height=6cm]{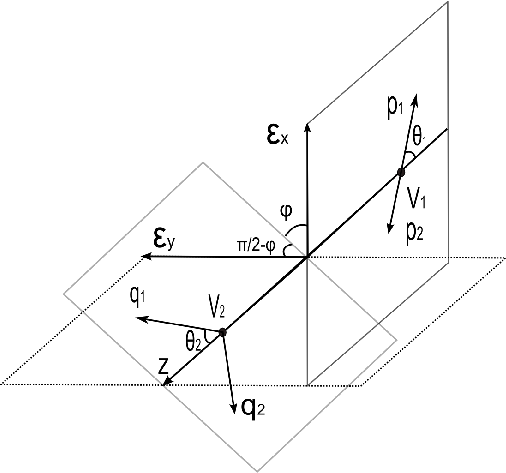}
\caption{} \label{etacvv-decay}
\end{figure}

According to the laws of angular momentum and parity conservation in
strong interaction we know that those two vector mesons are in the
state of total spin 1. From the Clebsch-Gordan decomposition, its
zero component in z-direction is
\begin{eqnarray}
\Psi_{|1,0\rangle} = \frac{1}{\sqrt{2}}( |1\rangle|-1\rangle -
|-1\rangle|1\rangle )_z\; , \label{spin1-z}
\end{eqnarray}
where the z-axis is chosen to be the moving direction of one vector
meson in another meson's rest frame, as shown in Figure
\ref{etacvv-decay}. Expressing the entangled state in bases
perpendicular to z-axis, one has
\begin{eqnarray}
\Psi_{|1,0\rangle} = \frac{1}{\sqrt{2}}(
|0\rangle_{\alpha_{\perp}}|0\rangle_{\alpha} -
|0\rangle_{\alpha}|0\rangle_{\alpha_{\perp}} )\; . \label{spin1-x}
\end{eqnarray}
Here, the subscript $\alpha$ indicates an arbitrary axis
perpendicular to z, and $\alpha_{\perp}$ means another base which is
perpendicular to both $z$ and $\alpha$ in terms of right-handed
system. The plane fixed by axes $\alpha$ and $\alpha_\perp$ is in
fact the $x$-$y$ plane, and $\alpha$ is the relative angle between
$x$- and $\alpha$-axes(or $y$- and $\alpha_\perp$-axes). Then one
can immediately get four different probabilities within the $x$-$y$
plane based on the QM, which are measurable quantities in
experiment. They are
\begin{eqnarray}
P(J_{\beta} = 0, J_{\alpha} = 0) &=&  \frac{1}{2}\sin^{2}(\alpha -
\beta) \; , \label{quant-constrain-1} \\P(J_{\beta} \neq 0,
J_{\gamma} = 0) & = & \frac{1}{2}\cos^2(\beta-\gamma) \; ,
\label{quant-constrain-2}
\\ P(J_{x} = 0, J_{\alpha} \neq 0) & = & \frac{1}{2}\cos^{2}\alpha \; ,
\label{quant-constrain-3} \\ P(J_{x} = 0, J_{\gamma} = 0) & = &
\frac{1}{2}\sin^2\gamma \; , \label{quant-constrain-4}
\end{eqnarray}
with
\begin{eqnarray}
&&P(J_{\beta} = 0, J_{\alpha} = 0) = |_{\beta}\langle
0|_{\alpha}\langle 0| \Psi\rangle_{|1,0\rangle}|^2 \; ,\nonumber \\
&&P(J_{\beta} \neq 0, J_{\gamma} = 0)  = \nonumber
\\
&&P( J_{\beta} = 1, J_{\gamma} = 0)  + P(J_{\beta} = -1, J_{\gamma}
= 0)\; , \nonumber \\
&&\hat{J}_{\alpha}  =  \hat{J}_{x} \cos\alpha  + \hat{J}_{y}
\sin\alpha \; .\nonumber
\end{eqnarray}
Here, $P(r_1,\; r_2)$ means the probability under the conditions of
$r_1$ and $r_2$, which are satisfied by the two partites of 1 and 2
of the entangled state, respectively. $|0\rangle_{\alpha}$ stands
for the eigenvector of angular momentum operator $\hat{J}_{\alpha}$
with eigenvalue $J_{\alpha}=0$ in the $x$-$y$ plane. From the
generalization of Hardy's argument to spin-$s$($s
=\frac{1}{2},1,\frac{3}{2},\cdots$) systems \cite{hardy-spin-s}, we
obtain the following LHVT constraint
\begin{eqnarray}
P(J_{x} = 0, J_{\gamma} = 0) & \leq & P(J_{x} = 0, J_{\alpha} \neq
0) \nonumber \\ &  + & P(J_{\beta} \neq 0, J_{\gamma} = 0) \nonumber \\
&  + & P(J_{\beta} = 0, J_{\alpha} = 0) \; .\label{LHVT-consntr}
\end{eqnarray}
Substituting the quantum mechanics results (\ref{quant-constrain-1})
- (\ref{quant-constrain-4}) into (\ref{LHVT-consntr}), one
immediately obtains
\begin{eqnarray}
\frac{1}{2} \sin^2{\gamma} \leq \frac{1}{2}[ \cos^{2}\alpha +
\cos^2(\beta-\gamma) + \sin^{2}(\alpha - \beta) ]\; .
\end{eqnarray}
This inequality is maximally violated by QM predictions while
$\alpha=3\pi/8,\beta = \pi/4,\gamma=5\pi/8$, which gives a
contradictive result of $\frac{2+\sqrt{2}}{8} \leq
\frac{6-3\sqrt{2}}{8}$.

In principle, in experiment one can determine the polarizations of
$\rho$s or $\phi$s through measuring the distributions of their
two-body exclusive decay products, the $\pi$s or $K$s, and then find
the difference between LR and QM predictions. Here, we find that
there exists another practical method to find out the difference in
the experiment, that is to perform the measurement on the quantities
in CH inequality event by event. Following we explain it in detail.

Suppose the probabilities of a count being triggered by the decays
of $V_{1},\; V_{2}$ polarizing along $\vec{n}_{1, 2}$ being
$p(\vec{n}_1,\lambda)$ and $q(\vec{n}_2,\lambda)$ respectively, then
we can readily get a CH inequality
\cite{Clauser-Horne,Junli-CF-Qiao}
\begin{eqnarray}
P(\vec{n}_{1},\vec{n}_{2}) - P(\vec{n}_{1},\vec{n}'_{2}) +
P(\vec{n}'_{1}, \vec{n}_{2}) + P(\vec{n}'_{1}, \vec{n}'_{2}) -
P(\vec{n}_1') -P(\vec{n}_2) \leq 0 \; . \label{ch-vv}
\end{eqnarray}
Here, $P(\vec{n}_1,\vec{n}_2) = \int p(\vec{n}_1,\lambda)
q(\vec{n}_2,\lambda) \rho(\lambda)\, \mathrm{d}\lambda $, and hence
$P(\vec{n}_1) = \int p(\vec{n}_1,\lambda) \rho(\lambda)\,
\mathrm{d}\lambda$ with $\lambda$ being a set of hidden variables.

In the quantum field theory, the differential decay width of $\eta_c
\to V_1(p,\vec{\epsilon}^{\, *})V_2(q,\vec{\epsilon}{\,'^*}) \to
P(p_1) P(p_2) P(q_1) P(q_2)$, as shown in Figure \ref{etacvv-decay},
takes the following form
\begin{eqnarray}
\frac{\mathrm{d}\Gamma_{\eta_c \rightarrow V_1V_2\rightarrow
\ldots}}{\mathrm{d}\varphi} & \propto & | \langle
\vec{n}_{1}|\langle \vec{n}_{2}|\Psi\rangle |^2
\label{prob-decay-width}
\end{eqnarray}
with
\begin{eqnarray}
|\langle \vec{n}_{1}|\langle \vec{n}_{2}|\Psi\rangle |^2
 & \equiv & P(\vec{n}_1, \vec{n}_2) \; , \label{qm-def-p}
\end{eqnarray}
where unit vectors $\vec{n}_{1}$ and $\vec{n}_{2}$ are normalized
projections of momenta $\vec{p}_1$ and $\vec{q}_1$ respectively of
the final pseudoscalar mesons in $x$-$y$ plane, and
\begin{eqnarray}
|\Psi\rangle  = \frac{1}{\sqrt{2}}(|\epsilon_{x}\rangle
|\epsilon'_{y}\rangle - |\epsilon_{y}\rangle |\epsilon'_{x}\rangle)
\label{waveform2}
\end{eqnarray}
with $\epsilon_{x,y}$ and $\epsilon'_{x,y}$ being the transverse
components of the polarization vectors $\vec{\epsilon}$ and
$\vec{\epsilon}\,'$. The wave function of (\ref{waveform2})
explicitly shows that the two vector mesons are entangled, the same
as (\ref{spin1-x}). Note that our concerned process is similar to
the $B$ to two-vector-meson process, the $B \to V_1 V_2$ \cite{bvv},
and hence the differential decay width (\ref{prob-decay-width}) can
be analogously obtained. The details of how to derive
(\ref{prob-decay-width}) and others will be presented elsewhere.

Given that the four final pseudoscalars move with momenta $p_1,\;
p_2,\; q_1$, and $q_2$, the azimuthal angle $\varphi$ between two
decay planes of the entangled vector meson pair then equals to the
angle between $\vec{n}_1$ and $\vec{n}_2$, as shown in Figure
\ref{etacvv-decay}. The magnitudes of $P(\vec{n}_1, \vec{n}_2)$ in
the CH inequality are therefore experimentally measurable, which is
obviously the probability density, up to an overall normalization
factor, from the definition of $P(\vec{n}_1, \vec{n}_2)$. That is
\begin{eqnarray}
P(\vec{n}_1, \vec{n}_2)  = \kappa \frac{N(\varphi + {\Delta\varphi})
- N(\varphi)}{N{\Delta\varphi}}\; . \label{exp-probability}
\end{eqnarray}
Here, $\kappa$ is a calculable normalization constant independent of
specific theory for our aim, $N$ is the total event number, and
$N(\varphi)$ is the event number within azimuthal angle $\varphi$,
the angle between $\vec{n}_1$ and $\vec{n}_2$. In above expression,
apart from the constant $\kappa$ the right-hand side is
experimentally measurable, i.e., the differential decay width of
$\eta_c$ to four pseudoscalar mesons divided by its total width via
intermediate vector meson $\phi$s, and is hence also theoretically
calculable in quantum theory. Inputting the theoretical and
experimental results into (\ref{ch-vv}), one may in principle find
the incompatibility of quantum theory with LR. However, in practice,
to perform the test of incompatibility the experiment efficiency
should be taken into account. The general inequality efficiency and
background levels was once discussed by Eberhard
\cite{Eberhard-ineq,hardy-effi}, and for the wave function
(\ref{waveform2}) the violation of inequality (\ref{ch-vv}) yields
the threshold efficiency $\eta > 82.8\%$ \cite{Brunner-Simon}.

To carry out the test of Bell type inequality, the decay angles
($\vec{n}_1,\vec{n}_2$) should generally be chosen actively by
experimenters, but this cannot be realized for mesons due to the
passive character of their decays \cite{Bertlmann-Hiesmayr}. Thus,
here only a restricted class of LR can be tested, like in
\cite{discuss-B-mesons}. A genuine Bell test also requires the decay
events of two vector mesons $V_1$ and $V_2$ to be space-like
separated. For the strongly decayed vector mesons ($\phi,\rho$,
etc.), we cannot guarantee for each particular event of $\eta_c \to
VV \to (PP)(PP)$ that the decays of two vector mesons are space-like
separated. The non-space-like decay events may induce the locality
loophole in experiment. Fortunately the ratio of the space-like
separated events can be determined by the magnitude of $\beta$
\cite{tornqvist}. And, for $\eta_c \to \phi\phi$, it is easy to find
that $\beta_{\phi} = 0.729$ \cite{Junli-CF-Qiao}, which is greater
than $0.59$, the lower bound required to test the LR
\cite{discuss-B-mesons}. Note that the $\beta$ has a negative effect
on the threshold counter efficiency $\eta$, and which then requires
a higher detection efficiency in experiment.

In summary, in this work we show that three photons in the decay of
ortho-positronium, or other similar onium, form a tripartite
entangled state, which is SLOCC equivalent to the GHZ state. Since
in experiment the ortho-positronium decays to three-photon process
can be well measured and huge data samples exist, new technology
development on the X-ray polarimeter may enable a direct measurement
on the hard photon polarization and hence find the difference
between quantum theory and LR. Due to the detection efficiency and
the unavoidable noise and losses, this cannot be considered as a
loophole-free experiment. Nevertheless, the present method appears
to be of broad interest because unlike the general production of GHZ
state in optical cases the non-locality of the three photons in
positronium decay arises in a dynamical process without
postselection \cite{hiemayr2}. The investigation of three partite
systems will show different features not only to that of bipartite
qubit systems, but bipartite qutrit and in general bipartite qudit
systems as well.

We find that two vector mesons in $\eta_c$ exclusive decay
automatically form a three-dimensional non-maximally entangled
state. A concrete scheme to test the LR in experiment via the
two-vector-meson entangled state is proposed. Since experimental
measurements on exclusive two body decay processes
\begin{eqnarray}
\rho\rightarrow \pi\pi,\; \phi\rightarrow K^+ K^- (K_L^0K^{0}_S)
\end{eqnarray}
are well-established, and give large branching fractions of
$\rho\rightarrow \pi\pi$ $\sim 100\%$ and $\phi\rightarrow K^+ K^-
(K_L^0K^{0}_S)$ $\sim 49.2\pm0.6\% \,(34.0\pm0.5\%)$ \cite{PDG}, we
believe that in current running experiments the CH type inequality
can be readily measured. It should be mentioned that the
non-space-like decay events may induce the so-called locality
loophole in experiment, which hinders the proposed tests to refute
the LR definitely. Nevertheless, the experimental realization of the
proposals in this work may extend the test on local realism into
high energy regime with multipartite and high dimension, which will
give us a more explicit conclusion in comparison with that from the
bipartite qubit results.

\vspace{.3cm} {\bf Acknowledgments} \vspace{.3cm}

This work was supported in part by the National Natural Science
Foundation of China(NSFC) under the grants 10491306, 10521003 and
10775179.

\newpage
\centerline{\bf\large Figure Captions}
\vspace{1.2cm}

\noindent {\bf Figure 1:} {The schematic Feynman diagram of
processes $\eta_c \rightarrow VV$. Here V stands for vector meson
$\phi$ or $\rho$.} \vspace{.5cm}

\noindent {\bf Figure 2:} {$\eta_c \to
V_1(p,\epsilon^*)V_2(q,\epsilon'^*)\to [P(p_1)P(p_2)]
[P(q_1)P(q_2)]$ decay kinematics in the rest frame of $V_1$. Here,
$V_1$, $V_2$ stand for vector meson pair of $\phi$s or $\rho$s,
$P(p_i)$ and $P(q_i)$ for the pseudoscalar mesons in $V_1$ and $V_2$
decays, the $K$s or $\pi$s\;.}

~\newpage

\end{document}